\documentclass[12pt]{iopart}
\usepackage{graphicx}
\newcommand{\ep}{\epsilon}
\newcommand{\SL}{S_\Lambda}
\newcommand{\SIL}{S_{I,\Lambda}}
\newcommand{\SFL}{S_{F,\Lambda}}
\newcommand{\vev}[1]{\left\langle #1 \right\rangle}
\newcommand{\K}[1]{K \left( #1/\Lambda \right)}
\newcommand{\Kz}[1]{K \left( #1/\Lambda_0 \right)}
\newcommand{\lb}{\left\lbrace}
\newcommand{\rb}{\right\rbrace}
\newcommand{\nn}{\nonumber}
\newcommand{\Op}{\mathcal{O}}
\newcommand{\N}{\mathcal{N}}
\newcommand{\PN}{[\Phi_N]}
\newcommand{\V}{\mathcal{V}}
\begin{document}

\rightline {\small KOBE-TH-09-10}

\title[ERG construction of the O(N) non-linear $\sigma$
model]{Perturbative construction of the two-dimensional O(N)
  non-linear sigma model with ERG} \author{B C
  L\"utf\"uo\u{g}lu\footnote{Present address: Department of Physics,
    Istanbul Technical University, Turkey} and H Sonoda}
\address{Physics Department, Kobe University, Kobe 657-8501, Japan}
\eads{\mailto{bcan@itu.edu.tr}, \mailto{hsonoda@kobe-u.ac.jp}}

\begin{abstract}
    We use the exact renormalization group (ERG) perturbatively to
    construct the Wilson action for the two-dimensional O(N) non-linear
    sigma model.  The construction amounts to regularization of a
    non-linear symmetry with a momentum cutoff.  A quadratically
    divergent potential is generated by the momentum cutoff, but its
    non-invariance is compensated by the jacobian of the non-linear
    symmetry transformation.
\end{abstract}

\maketitle

\section{Introduction\label{intro}}

The two dimensional O(N) non-linear $\sigma$ model is important for
its asymptotic freedom and dynamical generation of a mass gap.
Classically the model is defined by the action
\begin{equation}
S_{cl} = - \frac{1}{2 g} \int d^2 x\, \sum_{I=1}^N \partial_\mu
\Phi_I \partial_\mu \Phi_I 
\end{equation}
where the real scalar fields satisfy the non-linear constraint
$\sum_{I=1}^N \Phi_I \Phi_I = 1$.  Regarding the model as a classical
spin system, $g$ plays the role of the temperature; large $g$
encourages fluctuations of the fields, while small $g$ discourages
them.  The asymptotic freedom of the model, first shown in
\cite{Polyakov:1975rr}, implies not only the validity of perturbation
theory at short distances but also the generation of a mass gap due to
large field fluctuations at long distances.

The purpose of this paper is to apply the method of the exact
renormalization group (ERG) to renormalize the model consistently with
a momentum cutoff.  The perturbative renormalization of the model is
done usually with the dimensional regularization
\cite{Bollini:1972ui,'tHooft:1972fi}.  Its advantage is the manifest
O(N) invariance, but an external magnetic field (mass term) must be
introduced to avoid IR divergences.  Compared with the dimensional
regularization, the regularization with a momentum cutoff is
physically more appealing, but it is technically more complicated; the
O(N) invariance is not manifest, and a na\"ive sharp momentum cutoff,
inconsistent with shifts of loop momenta, cannot be used beyond
1-loop.

We can overcome the technical difficulties using the formulation of
field theory via ERG differential equations
\cite{Wilson:1973jj, Polchinski:1983gv}.  For a general
perturbative construction of theories with continuous symmetry, we
refer the reader to a recent review article \cite{Igarashi:2009tj},
and in this paper we give only the minimum background necessary for
our purposes.  ERG was first applied to the two dimensional O(N)
non-linear $\sigma$ model by Becchi \cite{Becchi:1996an}; we aim to
simplify and complete his analysis.  In particular, we give a
perturbative algorithm for constructing the Wilson action of the model
with a finite momentum cutoff $\Lambda$.  The Wilson action results
from an integration of fields with momenta larger than $\Lambda$, and
it is free from IR divergences without an external magnetic field.

Throughout the paper we use the Euclid metric and the following
notation for momentum integrals:
\begin{equation}
\int_p \equiv \int \frac{d^2 p}{(2\pi)^2}
\end{equation}
A short summary of this paper has appeared in sect.~6.4 of
\cite{Igarashi:2009tj}.

\section{Momentum cutoff}

We regularize the model using a UV momentum cutoff $\Lambda_0$.  The
bare action is given by
\begin{equation}
S_B = - \frac{1}{2} \int_p \frac{p^2}{K(p/\Lambda_0)} \phi_i (-p) \phi_i
 (p) + S_{I,B}
\label{bare action}
\end{equation}
where the subscript $i$, running from $1$ to $N-1$, is summed over.  The
interaction part is given by
\begin{eqnarray}
    S_{I,B} &=& \int d^2 x \, \left[ \Lambda_0^2\,  z_0
        \left(\phi^2/2\right)\right.\nn\\
&& \quad\left.  + z_1 \left(\phi^2/2\right) \,(- \partial^2) \frac{1}{2}
        \phi^2
        + z_2 \left(\phi^2/2\right) \, \phi_i (- \partial^2) \phi_i
        \,\right]
\end{eqnarray}
where we denote $\phi^2 = \phi_i\phi_i$.  $z_0, z_1, z_2$ are
functions of $\phi^2/2$ and depend logarithmically on the cutoff
$\Lambda_0$.  $S_{I,B}$ is the most general interaction action allowed
by the manifest O(N$-$1) invariance and perturbative renormalizability
in the absence of any dimensionful parameters.

The propagator given by the free part of (\ref{bare action}) is
proportional to the smooth cutoff function $K(p/\Lambda_0)$.  By
choosing $K(x)$ such that
\begin{enumerate}
\item $K(x)$ is a positive and non-increasing function of $x^2$,
\item $K(x) = 1$ for $x^2 < 1$,
\item $K(x)$ damps rapidly (faster than $1/x^2$) as $x^2 \to \infty$,
\end{enumerate}
we can regularize the UV divergences of the model.  

The renormalization functions $z_{0,1,2}$ must be fine tuned, first
for renormalizability, and then for the O(N) invariance.

\section{Wilson action}

The Wilson action with a finite momentum cutoff $\Lambda$ has two
parts:
\begin{equation}
\SL \equiv \SFL + \SIL
\end{equation}
The free part
\begin{equation}
\SFL \equiv - \frac{1}{2} \int_p \frac{p^2}{K(p/\Lambda)} \phi_i (-p) \phi_i
 (p)
\end{equation}
gives the propagator with a finite momentum cutoff $\Lambda$:
\begin{equation}
\vev{\phi_i (p) \phi_j (-p)}_{\SFL} = \delta_{ij} \frac{\K{p}}{p^2}
\end{equation}
The interaction part of the Wilson action is defined by
\begin{eqnarray}
    &&\exp \left[ \SIL [\phi] \right]
    \equiv \int [d\phi']\, \nn\\
    &&\quad \times \exp \left[ - \frac{1}{2} \int_p
        \frac{p^2}{\Kz{p} - \K{p}} \phi'_i (-p) \phi'_i (p) + S_{I,B}
        [\phi + \phi'] \right]\\
    && \qquad = \exp \left[ \frac{1}{2} \int_p
        \frac{\Kz{p}-\K{p}}{p^2} \frac{\delta^2}{\delta \phi_i (p)
          \delta \phi_i (-p)}\right] \cdot \exp \left[ S_{I,B} [\phi]
    \right]\nn
\end{eqnarray}
Alternatively, we can define $\SIL$ by the differential equation
\cite{Wilson:1973jj, Polchinski:1983gv}
\begin{equation}
- \Lambda \frac{\partial}{\partial \Lambda} \SIL = \frac{1}{2} \int_p
\frac{\Delta (p/\Lambda)}{p^2} 
\lb \frac{\delta \SIL}{\delta \phi_i (-p)} \frac{\delta \SIL}{\delta
  \phi_i (p)} + \frac{\delta^2 \SIL}{\delta \phi_i (-p) \delta \phi_i
  (p)} \rb
\label{diffeq}
\end{equation}
and the initial condition
\begin{equation}
\SIL\Big|_{\Lambda = \Lambda_0} = S_{I,B}
\label{initial}
\end{equation}

For a fixed $\Lambda$, we expand $\SIL$ up to two derivatives to
obtain
\begin{eqnarray}
    \SIL &=& \int d^2 x\, \left[ \Lambda^2 \, a \left( \ln
            \Lambda/\mu; \phi^2/2 
        \right) + A \left(\ln \Lambda/\mu; \phi^2/2\right)\,
        (- \partial^2) \frac{1}{2} 
        \phi^2 \right.\nn\\
    && \quad \left. + B \left(\ln \Lambda/\mu; \phi^2/2 \right)\,
        \phi_i (- \partial^2) \phi_i \,\right] \, + \cdots
\label{asymp}
\end{eqnarray}
where the dotted part contains four or more derivatives.  $a, A, B$
are functions of $\phi^2/2$, and they can be expanded as
\begin{equation}
\lb
\begin{array}{c@{~=~}l}
a \left( \ln \Lambda/\mu; \phi^2/2 \right) & \sum_{n=1}^\infty \frac{1}{n!}
    \left(\frac{\phi^2}{2}\right)^n \,
    a_n (\ln \Lambda/\mu) \\
A \left(\ln \Lambda/\mu; \phi^2/2\right)& \sum_{n=1}^\infty \frac{1}{n!} \left(
    \frac{\phi^2}{2} \right)^n\, A_n (\ln \Lambda/\mu)\\
B\left(\ln \Lambda/\mu; \phi^2/2\right)& \sum_{n=0}^\infty \frac{1}{n!} \left(
    \frac{\phi^2}{2} \right)^n\, B_n (\ln \Lambda/\mu)
\end{array}\right.
\end{equation}
The Taylor coefficients depend logarithmically on the cutoff
$\Lambda$.  We have chosen the ratio of $\Lambda$ to an arbitrary
renormalization scale $\mu$ as the argument of the logarithm.  The
initial condition (\ref{initial}) gives
\begin{equation}
\lb\begin{array}{c@{~=~}l}
a(\ln \Lambda_0/\mu; \phi^2/2) & z_0 (\phi^2/2)\\
A(\ln \Lambda_0/\mu; \phi^2/2) & z_1 (\phi^2/2)\\
B(\ln \Lambda_0/\mu; \phi^2/2) & z_2 (\phi^2/2)
\end{array}\right.
\end{equation}
The renormalization functions $z_{0,1,2}$ are determined so that
\begin{equation}
\lim_{\Lambda_0 \to \infty} \SIL
\label{continuum limit}
\end{equation}
exists for any finite $\Lambda$.  Using the BPHZ renormalization
scheme adapted to the Wilson action \cite{Pernici:1998tp,
  Sonoda:2002pb, Sonoda:2006ai}, we can choose $A(0; \phi^2/2)$ \&
$B(0; \phi^2/2)$ as any functions.  As will be explained in the next
section, the O(N) invariance constrains the choice of $A(0; \phi^2/2)$
\& $B(0; \phi^2/2)$.

Alternatively, we can construct the continuum limit (\ref{continuum
  limit}) directly without starting from a bare action.  We demand
that the dotted part of (\ref{asymp}) is multiplied by the inverse
powers of $\Lambda$.  For given $A(0; \phi^2/2)$ \& $B(0; \phi^2/2)$,
the differential equation (\ref{diffeq}) uniquely determines $a \left(
    \ln \Lambda/\mu; \phi^2/2\right)$ and the dotted part of
(\ref{asymp}) \cite{Pernici:1998tp, Sonoda:2002pb, Sonoda:2006ai}.
This is the preferred approach we adopt in the rest of the paper.  In
\ref{basic}, we summarize the basic properties of the correlation
functions calculated with $\SL$.

\section{WT identity for O(N)\label{WT}}

The Wilson action is determined uniquely in terms of $A(0; \phi^2/2)$
\& $B(0; \phi^2/2)$.  For the O(N) symmetry, we must choose $A(0;
\phi^2/2)$ \& $B(0; \phi^2/2)$ appropriately.  In this and the
following two sections, we aim to complete the analysis of Becchi
given in sect.~6 of \cite{Becchi:1996an}.

The Wilson action has manifest O(N$-$1) invariance.  To insure the
full O(N) invariance, we must demand the invariance of the action
under the following infinitesimal transformation:
\begin{equation}
\delta \phi_i (p) = \K{p} \ep_i \PN (p)
\label{deltaphi}
\end{equation}
where $\ep_i$ is an infinitesimal constant, and $\PN$ is the
composite operator for the $N$-th component of the O(N) vector, whose
$i$-th component is proportional to $\phi_i$.  More precisely, $\PN$
is defined by the ERG differential equation
\begin{eqnarray}
&&- \Lambda \frac{\partial}{\partial \Lambda} \PN (p)\nn\\
&&\quad = \int_q \frac{\Delta (q/\Lambda)}{q^2} \lb \frac{\delta \SIL}{\delta
  \phi_i (q)} \frac{\delta}{\delta \phi_i (q)} + \frac{1}{2}
\frac{\delta^2}{\delta \phi_i (q) \delta \phi_i (-q)} \rb \PN
(p)\label{PhiNdiffeq}
\end{eqnarray}
and the derivative expansion
\begin{equation}
\int_p \e^{i p x} \PN (p) = P \left(\ln \Lambda/\mu; \phi
(x)^2/2\right) + \cdots
\end{equation}
where the dotted part, proportional to the inverse powers of
$\Lambda$, contains derivatives of $\phi_i (x)$. $\PN$ is
parameterized by a function
\begin{equation}
P\left(0; \phi^2/2\right)
\end{equation}
which is arbitrary as far as perturbative renormalizability of
$\PN$ is concerned.  (The composite operators, the concept of
which was first introduced in sect.~5 of \cite{Becchi:1996an}, can be
considered as infinitesimal deformations of the Wilson action, and
they satisfy the same linear ERG differential equation as
(\ref{PhiNdiffeq}).  A composite operator vanishes identically if the
leading part in the derivative expansion, the part multiplied by the
non-negative powers of $\Lambda$, vanishes.  For more details, see
sect.~4 of \cite{Igarashi:2009tj}.)

Following Becchi \cite{Becchi:1996an}, we now define the
Ward-Takahashi (WT) composite operator for (\ref{deltaphi}) by
\begin{eqnarray}
\Sigma_\Lambda &\equiv& \int_p \left[ \frac{\delta \SL}{\delta \phi_i (p)}
  \delta \phi_i (p) + \frac{\delta}{\delta \phi_i (p)}
  \delta \phi_i (p) \right] \nn\\
&=& \ep_i \int_p \K{p}  \left[ \frac{\delta \SL}{\delta \phi_i (p)}
  \PN (p) + \frac{\delta \PN (p)}{\delta \phi_i (p)}
  \right] \label{WToperator}
\end{eqnarray}
This satisfies the same ERG linear differential equation as
(\ref{PhiNdiffeq}).  The WT identity
\begin{equation}
\Sigma_\Lambda = 0
\label{WT-WT}
\end{equation}
is the ``quantum'' invariance of the Wilson action under
(\ref{deltaphi}), whereby the non-trivial jacobian of (\ref{deltaphi})
is taken into account.  Concrete loop calculations show that the
coefficient function $a (\ln \Lambda/\mu; \phi^2/2)$, corresponding to
the quadratically divergent potential in the bare action, is
non-vanishing.  But its non-invariance under (\ref{deltaphi}) is cancelled
by the jacobian.  

Taking the correlation of $\Sigma_\Lambda$ with the elementary fields,
we obtain the usual WT identity from (\ref{WT-WT}):
\begin{equation}
\sum_{j=1}^n \ep_{i_j} \vev{\phi_{i_1} (p_1) \cdots \Phi_N (p_j) \cdots
  \phi_{i_n} (p_n)}^\infty = 0
\end{equation}
where the renormalized correlation function
\begin{eqnarray}
&&\vev{\phi_{i_1} (p_1) \cdots \Phi_N (p_j) \cdots
  \phi_{i_n} (p_n)}^\infty \nn\\
&\equiv& \prod_{k \ne j} \frac{1}{\K{p_k}}
\,\times \vev{\phi_{i_1} (p_1) \cdots \PN (p_j) \cdots \phi_{i_n}
  (p_n)}_{\SL}
\end{eqnarray}
is independent of the cutoff $\Lambda$.  (This $\Lambda$ independence
is a consequence of the differential equations (\ref{diffeq},
\ref{PhiNdiffeq}).  See, for example, sect.\,4.1 of
\cite{Igarashi:2009tj} for more explanations.) 

For $\SL$ to satisfy (\ref{WT-WT}), we must fine tune not only
$A(0;\phi^2/2)$ \& $B(0;\phi^2/2)$ but also $P(0; \phi^2/2)$.  In the
next two sections, we will show the possibility of such fine tuning.

\section{Tree level\label{tree}}

We expand $\SL, \SIL$, etc., in the number of loops.  We use a
superscript ${}^{(l)}$ to denote the $l$-loop level:
\begin{equation}
\begin{array}{l}
\SL = \sum_{l=0}^\infty \SL^{(l)},\quad
\PN (p) = \sum_{l=0}^\infty \PN^{(l)} (p),\quad
\Sigma_\Lambda = \sum_{l=0}^\infty \Sigma_\Lambda^{(l)},\\
A\left(\ln \Lambda/\mu; \phi^2/2\right) = \sum_{l=0}^\infty A^{(l)} \left(\ln
\Lambda/\mu; \phi^2/2\right),\,\cdots\\
A_i \left(\ln \Lambda/\mu\right) = \sum_{l=0}^\infty A_i^{(l)}
\left(\ln \Lambda/\mu\right),\,\cdots\\
A\left(0; \phi^2/2\right) = \sum_{l=0}^\infty A^{(l)}
\left(\phi^2/2\right),\,\cdots
\end{array}
\end{equation}
In this section, we show how to tune
\begin{equation}
    A^{(0)} \left(\phi^2/2\right),\quad B^{(0)}
    \left(\phi^2/2\right),\quad P^{(0)} \left(\phi^2/2\right) 
\end{equation}
to satisfy $\Sigma^{(0)}_\Lambda = 0$.

The leading part of the derivative expansion of $\SL^{(0)}$ is given
by the classical action:
\begin{equation}
\SL^{(0)} = S_{cl} + \cdots
\end{equation}
$S_{cl}$ is independent of $\Lambda$, and we can write
\begin{eqnarray}
S_{cl} &=& \int d^2 x\, \Big[ - \frac{1}{2} \partial_\mu
    \phi_i \partial_\mu \phi_i\nn\\
&&\quad +  A^{(0)} \left(\phi^2/2\right) (- \partial^2)
  \frac{\phi^2}{2} + B^{(0)} \left(\phi^2/2\right) \phi_i (- \partial^2) \phi_i
\Bigg]\label{tree-S} 
\end{eqnarray}
Likewise, the derivative expansion of
$\PN^{(0)}$ gives
\begin{equation}
\int_p \e^{ipx} \PN^{(0)} (p) = P^{(0)} \left(\phi^2/2\right) + \cdots
\label{tree-P}
\end{equation}
As a convention, we can choose
\begin{equation}
A_0^{(0)} = B_0^{(0)} = 0,\quad P_0^{(0)} = 1
\label{tree-normalization}
\end{equation}

At tree level, the WT identity gives
\begin{equation}
\Sigma_\Lambda^{(0)} \equiv \ep_i \int_p \K{p} \frac{\delta
  \SL^{(0)}}{\delta \phi_i (p)} \PN^{(0)} (p) = 0
\end{equation}
The derivative expansion gives
\begin{equation}
\Sigma_{cl}\equiv \ep_i \int d^2 x \,  \frac{\delta S_{cl}}{\delta
  \phi_i (x)} P^{(0)} \left(\phi (x)^2/2\right) = 0\label{tree-WT}
\end{equation}
Substituting (\ref{tree-S}) into the above, we obtain
\begin{eqnarray}
&&\Sigma_{cl} = \ep_i \int d^2 x\,  \phi_i
\Big[ \partial_\mu \phi_j \partial_\mu \phi_j \lb P^{(0)'} - (2
A^{(0)'} + B^{(0)'}) P^{(0)} - 2 P^{(0)'} B^{(0)} \rb\nn\\
&&\, + \phi_j \partial^2 \phi_j \lb P^{(0)'} - 2 (A^{(0)'} +
 B^{(0)'}) P^{(0)} - 2 P^{(0)'} B^{(0)} \rb\\
&&\, + (\phi_j \partial_\mu \phi_j)^2 \lb (1 - 2 B^{(0)})
P^{(0)''} - (A^{(0)''} + B^{(0)''}) P^{(0)}  - 2 B^{(0)'}
P^{(0)'} \rb \Big]\nn
\end{eqnarray}
where the prime denotes a derivative with respect to $\phi^2/2$.  For
$\Sigma_{cl}$ to vanish, we must satisfy the following three
equations: \numparts
\begin{eqnarray}
&&(1 - 2 B^{(0)}) P^{(0)'} - (2 A^{(0)'} + B^{(0)'}) P^{(0)} = 0
\label{tree1}\\
&&(1 - 2 B^{(0)}) P^{(0)'} - 2 (A^{(0)'} + B^{(0)'}) P^{(0)} = 0
\label{tree2}\\
&&(1 - 2 B^{(0)}) P^{(0)''} - 2 B^{(0)'} P^{(0)'} - (A^{(0)''} +
B^{(0)''}) P^{(0)} = 0
\label{tree3}
\end{eqnarray}
\endnumparts

From (\ref{tree1}) and (\ref{tree2}), we get
\begin{equation}
B^{(0)'} (x) P^{(0)} (x) = 0
\end{equation}
where we write $x \equiv \phi^2/2$ for short.  Since $P^{(0)} (x) \ne
0$, we obtain $B^{(0)'} (x) = 0$; hence using
(\ref{tree-normalization}) we obtain
\begin{equation}
B^{(0)} (x) = 0
\end{equation}
Thus, (\ref{tree1}) gives
\begin{equation}
P^{(0)'} (x) = 2 A^{(0)'} (x) P^{(0)} (x)
\end{equation}
Using (\ref{tree-normalization}), we obtain
\begin{equation}
P^{(0)} (x) = \exp \left[ 2 A^{(0)} (x) \right] 
\end{equation}
Finally, (\ref{tree3}) gives
\begin{equation}
P^{(0)''} (x) = A^{(0)''} (x) P^{(0)} (x)
\end{equation}
This is solved by
\begin{equation}
A^{(0)} (x) = \frac{1}{4} \ln \left( 1 - 2 c \,x \right)
\end{equation}
where $c$ is an arbitrary constant. Hence, we obtain
\begin{equation}
P^{(0)} (x) = \sqrt{ 1 - 2 c x}
\end{equation}

The constant $c$ may be chosen either positive or negative.
If we choose a positive $c = g > 0$, then we obtain
\begin{equation}
\Phi_N = \sqrt{ 1 - g \phi^2}
\end{equation}
appropriate for the classical O(N) non-linear $\sigma$ model.  If we
choose a negative $c = - g < 0$ instead, we obtain
\begin{equation}
\Phi_N = \sqrt{1 + g \phi^2}
\end{equation}
appropriate for the classical O(N$-$1,1) non-linear $\sigma$ model.
We make the first choice.

To summarize, we have obtained
\begin{equation}
\lb\begin{array}{c@{~=~}l}
A^{(0)} (x) & \frac{1}{4} \ln (1 - 2 g x)\\
B^{(0)} (x) & 0\\
P^{(0)} (x) & \sqrt{ 1 - 2 g x}
\end{array}\right.
\label{tree-results}
\end{equation}
where $g$ is an arbitrary \textbf{positive} coupling constant.  The
corresponding classical action is given by the familiar expression
\begin{equation}
S_{cl} = - \frac{1}{2 g} \int d^2 x\, \left[ g \partial_\mu
    \phi_i \partial_\mu \phi_i + \partial_\mu \sqrt{1 - g
      \phi^2} \cdot \partial_\mu \sqrt{1 - g \phi^2} \right]
\end{equation}

\section{Loop levels\label{loop}}

Let us now assume that we have determined $\SL$ and $\PN$ up to
$l$-loop level ($l \ge 0$) such that
\begin{equation}
\Sigma_\Lambda^{(0)}  = \cdots = \Sigma_\Lambda^{(l)} = 0
\label{hypothesis}
\end{equation}
Under this induction hypothesis, we wish to determine $\SL^{(l+1)}$
and $\PN^{(l+1)}$ (or equivalently $A^{(l+1)}(x), B^{(l+1)} (x),
P^{(l+1)} (x)$) such that
\begin{equation}
\Sigma_\Lambda^{(l+1)} = 0
\end{equation}
Note that $\Sigma_\Lambda$ is a composite operator, satisfying the
same ERG differential equation as (\ref{PhiNdiffeq}).  Applying
the loop expansion and using the induction hypothesis, we find
\begin{equation}
    - \Lambda \frac{\partial}{\partial \Lambda} \Sigma_\Lambda^{(l+1)}
    = \int_p \frac{\Delta (p/\Lambda)}{p^2} \frac{\delta
      \SIL^{(0)}}{\delta \phi_i (-p)} \frac{\delta
      \Sigma_\Lambda^{(l+1)}}{\delta \phi_i (p)} 
\end{equation}
Calling the leading part of the derivative expansion of
$\Sigma_\Lambda^{(l+1)}$ by $\tilde{\Sigma}^{(l+1)}$, we obtain
\begin{equation}
- \Lambda \frac{\partial}{\partial \Lambda} \tilde{\Sigma}^{(l+1)} = 0
\end{equation}
since $\Delta (p/\Lambda) = 0$ for $p^2 < \Lambda^2$.
Hence, $\tilde{\Sigma}^{(l+1)}$ is independent of $\Lambda$.
Thus, we obtain
\begin{eqnarray}
\tilde{\Sigma}^{(l+1)} [\phi] &=& \ep_i
\int d^2 x\,  \phi_i
\left[ \partial_\mu
    \phi_j \partial_\mu \phi_j \cdot s_1 \left(\phi^2/2\right) \right.\nn\\
&& \left.\, + \phi_j \partial^2 \phi_j \cdot s_2 \left(\phi^2/2\right)
+ (\phi_j \partial_\mu \phi_j)^2 \cdot s_3 \left(\phi^2/2\right) \right]
\end{eqnarray}
where $s_i (\phi^2/2)\,(i=1,2,3)$ are functions of $\phi^2/2$,
independent of $\ln \Lambda/\mu$.  

The definition (\ref{WT-WT}) of $\Sigma_\Lambda$ gives the
decomposition
\begin{equation}
\Sigma_\Lambda^{(l+1)} = \Sigma_\Lambda^{(l+1),t} + \Sigma_\Lambda^{(l+1),u}
\end{equation}
where
\numparts
\begin{eqnarray}
\Sigma_\Lambda^{(l+1),t} &=& \ep_i \int_p \K{p} \left[ \frac{\delta
  \SL^{(l+1)}}{\delta \phi_i (p)} \PN^{(0)} (p) +
\frac{\delta \SL^{(0)}}{\delta \phi_i (p)} \PN^{(l+1)} (p)\right]\\
\Sigma_\Lambda^{(l+1),u} &=& 
\ep_i \int_p \K{p} \left[ \sum_{k=1}^{l} \frac{\delta
  \SL^{(k)}}{\delta \phi_i (p)} \PN^{(l+1-k)} (p)
+ \frac{\delta \PN^{(l)} (p)}{\delta \phi_i (p)} \right]
\end{eqnarray}
\endnumparts
Only $\Sigma_\Lambda^{(l+1),t}$ depends on 
$A^{(l+1)} (x), B^{(l+1)} (x), P^{(l+1)} (x)$, and
$\Sigma_\Lambda^{(l+1),u}$ are determined by $\SL$ and $\PN$ up to
$l$-loop. 

Therefore, the functions $s_i (x)$ are given as the sum
\begin{equation}
s_i (x) = t_i (x) + u_i (x)\quad (i=1,2,3)
\end{equation}
where $t_i (x)$ are linear in $A^{(l+1)} (x), B^{(l+1)} (x), P^{(l+1)} (x)$,
and $u_i (x)$ are determined by the lower loop functions.  We obtain
explicitly
\numparts
\begin{eqnarray}
    t_1 (x) &=& P^{(l+1)'} - (2 A^{(l+1)'}+B^{(l+1)'}) P^{(0)}\nn\\
    &&\quad - 2 A^{(0)'} P^{(l+1)} - 2 P^{(0)'} B^{(l+1)}\\
    t_2 (x) &=& P^{(l+1)'} - 2 (A^{(l+1)'}+B^{(l+1)'})
    P^{(0)}\nn\\
    &&\quad  -  2 A^{(0)'}
    P^{(l+1)} - 2 P^{(0)'} B^{(l+1)} \\
    t_3 (x) &=& P^{(l+1)''} - (A^{(l+1)''} + B^{(l+1)''})
    P^{(0)}\nn\\ &&\quad -
    A^{(0)''} P^{(l+1)} - 2 P^{(0)''} B^{(l+1)}
    - 2 B^{(l+1)'} P^{(0)'}
\end{eqnarray}
\endnumparts
There is no relation among the $t (x)$'s.  Thus, whatever $u (x)$'s
are, we can solve the equations
\begin{equation}
s_i (x) = t_i (x) + u_i (x) = 0\quad (i=1,2,3)
\end{equation}
Using (\ref{tree-results}), the solution is obtained explicitly as
follows: \numparts
\begin{eqnarray}
&&B^{(l+1)} (x) = B^{(l+1)} (0) + \int_0^x dy \frac{- u_1 (y) + u_2
  (y)}{\sqrt{1 - 2 g y}}\\ 
&&\frac{d}{dx} A^{(l+1)} (x) = \frac{1}{(1 - 2 g x)^2} \Bigg[
\,A^{(l+1)'} (0) \nn\\
&&\quad + \int_0^x dy \lb - 2 g^2 B^{(l+1)} (y) + (1-2 g y) B^{(l+1)''}
    (y)\right.\nn\\
&&\qquad + g \sqrt{1 - 2 g y} \left( - 2 u_1 (y) + u_2 (y) \right)\nn\\
&&\qquad \left.+ (1-2 g y)^{\frac{3}{2}} \left( 2 u'_1 (y) - u'_2 (y) - u_3
    (y) \right) \rb\,\Bigg]\\
&&P^{(l+1)} (x) = \sqrt{1 - 2 g x} 
\bigg[ P^{(l+1)} (0) + \int_0^x dy\, \bigg\lbrace
2 A^{(l+1)'} (y) \nn\\
&&\qquad\qquad\qquad - \frac{2 g}{1 - 2 g y} B^{(l+1)}
    (y) + \frac{- 2 u_1 (y) + u_2 (y)}{\sqrt{1 - 2 g y}} \bigg\rbrace \bigg] 
\end{eqnarray}
\endnumparts
Note that
\begin{equation}
A^{(l+1)'} (0)\,,\quad
B^{(l+1)} (0) \,,\quad
P^{(l+1)} (0)
\end{equation}
are left undetermined as constants of integration.  This is expected,
since $A^{(l+1)'} (0)$ normalizes the coupling $g$, $B^{(l+1)} (0)$
normalizes the field $\phi^i$, and $P^{(l+1)} (0)$ normalizes the
composite operator $\PN$.  For example, we can adopt the convention
\cite{Pernici:1998tp, Sonoda:2002pb, Sonoda:2006ai}
\numparts
\begin{eqnarray}
A_1 \left(\ln \Lambda/\mu\right) \Big|_{\Lambda=\mu} &=&
\frac{\partial}{\partial x} A (0;x)\Big|_{x=0} = - \frac{g}{2}
\label{normA1}\\
B_0 \left(\ln \Lambda/\mu\right)\Big|_{\Lambda=\mu} &=& B(0;0) = 0
\label{normB0}\\
P_0 \left(\ln \Lambda/\mu\right)\Big|_{\Lambda=\mu} &=& P(0;0) = 1
\label{normP0}
\end{eqnarray}
\endnumparts
analogous to the minimal subtraction for dimensional regularization
\cite{'tHooft:1973mm}.  This concludes our inductive construction of
the O(N) non-linear $\sigma$ model.

\section{1-loop results}

Let us give explicitly the 1-loop corrections to 2- and 4-point
vertices in the Wilson action.  (A little more details are given in
\ref{loop-calculations}.)  For the 2-point vertex, we find \numparts
\begin{eqnarray}
a_1^{(1)} &=& \frac{g}{2} \int_q \Delta (q) = g \int_q K(q)\\
B_0^{(1)} &=& \frac{g}{4\pi} \ln \Lambda/\mu\label{B01}
\end{eqnarray}
\endnumparts
and for the 4-point vertex, we find
\numparts
\begin{eqnarray}
a_2^{(1)} &=& 2 g^2 \int_q \Delta (q) K(q) = 2 g^2 \int_q K(q)^2\\
A_1^{(1)} &=& N \frac{g^2}{4 \pi} \ln \Lambda/\mu\label{A11}\\
B_1^{(1)} &=& \mathrm{const}
\end{eqnarray}
\endnumparts
We also find the 2- and 4-point vertices for the composite operator
$\PN (p)$: 
\numparts
\begin{eqnarray}
P_0^{(1)} &=& \frac{N-1}{4 \pi} g \ln \Lambda/\mu\\
P_1^{(1)} &=& (N-1) \frac{g^2}{4\pi} \ln \Lambda/\mu + \mathrm{const}
\end{eqnarray}
\endnumparts
We have fixed the $\Lambda$ independent part of $A_1^{(1)}$,
$B_0^{(1)}$, and $P_0^{(1)}$ using the convention
(\ref{normA1}, \ref{normB0}, \ref{normP0}).

The two constants in $B_1^{(1)}$ and $P_1^{(1)}$ are left undetermined by
the ERG differential equations.  They are determined 
by the WT identity.  Calculating 
\begin{equation}
\ep_i \int_p \K{p} \frac{\delta \PN^{(0)} (p)}{\delta \phi_i (p)}
\end{equation}
only up to cubic in fields and up to two derivatives, we obtain
\numparts
\begin{eqnarray}
u_1^{(1)} (0) &=& - g^2 \left( \int_q \frac{K(q)(1-K(q))}{q^2} +
    \frac{1}{4\pi}\right)\\  
u_2^{(1)} (0) &=& g^2 \int_q \frac{K(q)}{q^2} \left( \frac{1}{4}
    \tilde{\Delta} (q) - 2 (1-K(q)) \right)
\end{eqnarray}
\endnumparts
where
\begin{equation}
\tilde{\Delta} (q) \equiv - 2 q^2 \frac{d}{dq^2} \Delta (q)
\end{equation}
Hence, we obtain
\numparts
\begin{eqnarray}
B_1^{(1)} &=& g^2 \left( \frac{1}{4\pi} - \int_q
    \frac{K(1-K)}{q^2} + \frac{1}{4} \int_q \frac{K
      \tilde{\Delta}}{q^2} \right)\\
P_1^{(1)}\Big|_{\Lambda=\mu} &=& g^2 \left( \frac{1}{2\pi} + \frac{1}{4}
    \int_q \frac{K \tilde{\Delta}}{q^2}\right)
\end{eqnarray}
\endnumparts

In \ref{beta} we explain how to obtain the beta function of $g$ and
anomalous dimension of $\phi_i$ in the ERG approach.  The above 1-loop
results reproduce the well known results first obtained in
\cite{Polyakov:1975rr}:
\begin{equation}
    \beta (g) \simeq (N-2) \frac{g^2}{2 \pi},\quad \gamma (g) \simeq \frac{g}{4
      \pi}\label{1loop-betagamma}
\end{equation}

\section{Concluding remarks}

In this paper we have applied the ERG formulation of quantum field
theory for the perturbative construction of the two-dimensional
non-linear $\sigma$ model.  The model is parameterized by three
renormalization functions $A (0;x), B(0;x), P(0;x)$, and we have shown
how to tune these by imposing the WT identity (\ref{WT-WT}).  Only
short-distance physics can be explored perturbatively, and
long-distance physics needs non-trivial approximations, such as $1/N$.
For the $1/N$ expansions it is common to linearize the O(N) symmetry
using an auxiliary field; it would be interesting to extend the ERG
formulation to accommodate the auxiliary field.

\appendix

\section{Basic properties of the correlation functions\label{basic}}

The correlation functions of a Wilson action $\SL$ are dependent on
the cutoff $\Lambda$.  Using the inverse of the cutoff function,
however, we can easily construct $\Lambda$ independent correlation
functions: 
\begin{eqnarray}
\vev{\phi_i (p) \phi_j (-p)}^\infty &\equiv& \frac{1}{\K{p}^2}
\vev{\phi_i (p) \phi_j (-p)}_{\SL} \nn\\
&&\quad + \delta_{ij} \frac{1 - 1/\K{p}}{p^2}\\
\vev{\phi_{i_1} (p_1) \cdots \phi_{i_n} (p_n)}^\infty &\equiv&
\prod_{j=1}^n \frac{1}{\K{p_j}} \cdot \vev{\phi_{i_1} (p_1) \cdots
  \phi_{i_n} (p_n)}_{\SL}
\end{eqnarray}
for $n \ge 4$.  The $\Lambda$ independence of these correlation
functions is a consequence of the ERG differential equation
(\ref{diffeq}).  See sect.~2 of \cite{Igarashi:2009tj} for more
details.

Similarly, given a composite operator $\Op_\Lambda$ that satisfies the
same linear ERG differential equation as (\ref{PhiNdiffeq}), we can
construct $\Lambda$ independent correlation functions by
\begin{equation}
    \vev{\Op \, \phi_{i_1} (p_1) \cdots \phi_{i_n} (p_n)}^\infty \equiv
    \prod_{j=1}^n \frac{1}{\K{p_j}} \cdot \vev{\Op_\Lambda\,
      \phi_{i_1} (p_1) \cdots \phi_{i_n} (p_n)}_{\SL}
\end{equation}
See sect.~4 of \cite{Igarashi:2009tj} for more details.

\section{1-loop calculations\label{loop-calculations}}

The interaction part of the classical action is given by
\begin{eqnarray}
S_{I,cl} &\equiv&
\int d^2 x (-\partial^2) \frac{\phi^2}{2} \cdot \frac{1}{4} \ln
\left(1 - g \phi^2 \right)\nn\\
&=& \int d^2 x (-\partial^2) \frac{\phi^2}{2} \cdot \frac{-1}{4}
\sum_{n=1}^\infty (2g)^n (n-1)! \cdot \frac{1}{n!}
\left(\frac{\phi^2}{2}\right)^n 
\end{eqnarray}
Hence,
\begin{equation}
A_n^{(0)} = - \frac{1}{4} (n-1)! (2 g)^n
\end{equation}
\begin{figure}
\begin{center}
\includegraphics{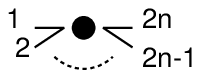}
\caption{Tree level vertex ($n \ge 2$)}
\label{vertex}
\end{center}
\end{figure}
Thus, for the graph in \Fref{vertex}, we obtain the
Feynman rule
\begin{equation}
\delta_{i_1 i_2} \cdots
\delta_{i_{2n-1} i_{2n}} \lb (p_1+p_2)^2 + \cdots + (p_{2n-1} +
  p_{2n})^2 \rb A_{n-1}^{(0)}
\end{equation}

\begin{figure}
\begin{center}
\includegraphics{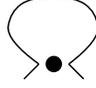}
\end{center}
\caption{$\V_2^{(1)} (p)$: 1-loop correction to the two-point vertex
  with momentum $p$}
\label{twopoint1loop}
\end{figure}

As the simplest example, we consider the 1-loop contribution to the
two-point vertex given by the Feynman graph in \Fref{twopoint1loop}.
The ERG differential equation gives
\begin{eqnarray}
- \Lambda \frac{\partial}{\partial \Lambda} \V_2^{(1)} (p) &=&
\int_q \frac{\Delta (q/\Lambda)}{q^2} 2 (p+q)^2 A_1^{(0)} 
= - \frac{g}{2} \int_q \frac{\Delta (q/\Lambda)}{q^2} 2 (p+q)^2\nn\\
&=& - g \int_q \frac{\Delta (q/\Lambda)}{q^2} (p^2 + q^2)
= - g \left[ \Lambda^2 \int_q \Delta (q) + \frac{1}{2\pi} p^2
\right]
\end{eqnarray}
where we have used
\begin{equation}
\int_q \frac{\Delta (q)}{q^2} = \frac{1}{2\pi}
\end{equation}
Hence, integrating this over $\Lambda$, we obtain
\begin{equation}
\V_2^{(1)} (p) = \Lambda^2 \frac{g}{2} \int_q \Delta (q) +
2 p^2 \frac{g}{4 \pi} \ln \Lambda/\mu 
\end{equation}
This gives
\begin{equation}
    a_1^{(1)} = \frac{g}{2} \int_q \Delta (q),\quad
    B_0^{(1)} \left(\ln \Lambda/\mu\right)  = \frac{g}{4 \pi} \ln
    \Lambda/\mu
\end{equation}
where we have used the normalization condition $B_0^{(1)} (0) = 0$.

\begin{figure}
\begin{center}
\includegraphics{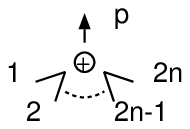}
\caption{$2n$-point vertex for $\PN (p)$}
\label{PhiNvertex}
\end{center}
\end{figure}

\begin{figure}
\begin{center}
\includegraphics{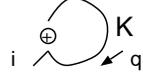}
\caption{1-loop contribution to the 1-point vertex of the jacobian 
(\ref{1loop-jacobian})}
\label{PhiN1pt}
\end{center}
\end{figure}

As another example, let us consider the 1-loop contribution to the
1-point vertex of the jacobian:
\begin{equation}
\int_q \K{q} \frac{\delta \PN^{(0)} (q)}{\delta \phi_i (q)}
\label{1loop-jacobian}
\end{equation}
Now, the leading part of the derivative expansion of $\PN^{(0)}$ is
given by
\begin{equation}
\int_p \e^{i p x} \PN^{(0)} (p) = P^{(0)} \left(\phi (x)^2/2\right)
+ \cdots
\end{equation}
where
\begin{equation}
P^{(0)} (x) = \sqrt{1 - 2 g x} = 1 - \sum_{n=1}^\infty
\frac{(2n-2)!}{2^{n-1} (n-1)!}\, g^n \cdot \frac{x^n}{n!}
\end{equation}
Hence, we obtain
\begin{equation}
P_0^{(0)} = 1,\quad
P_{n\ge 1}^{(0)} = - \,\frac{(2n-2)!}{2^{n-1} (n-1)!}\, g^n
\end{equation}
Let us denote the $2n$-point vertex $P_n^{(0)}$ for $\PN$ by
\Fref{PhiNvertex}.  Then, the one-point vertex for the 1-loop
jacobian is given by \Fref{PhiN1pt} and calculated as
\begin{equation}
    \int_q \K{q} P^{(0)}_1 = - \Lambda^2 \,g  \int_q K(q) 
= - \Lambda^2 \, \frac{g}{2} \int_q \Delta (q)
\end{equation}
This cancels the contribution of the $a_1^{(1)}$ term to
$\Sigma_\Lambda^{(1)}$.

\section{Beta function and anomalous dimension\label{beta}}

The derivation of the mass independent beta functions and anomalous
dimensions in the ERG formalism has been discussed in
\cite{Pernici:1998tp} and \cite{Sonoda:2006ai}.

\subsection{$\mu$ dependence of the Wilson action}

The Wilson action $\SL$ for a different choice of
$\mu$ satisfies the same ERG differential equation (\ref{diffeq}).
Hence, 
\begin{equation}
\Psi_\Lambda \equiv - \mu \partial_\mu \SL
\end{equation}
is a composite operator satisfying the ERG differential equation
\begin{equation}
- \Lambda \frac{\partial}{\partial \Lambda} \Psi_\Lambda =
\int_p \frac{\Delta (p/\Lambda)}{p^2} \lb
\frac{\delta \SIL}{\delta \phi_i (p)} \frac{\delta \Psi_\Lambda}{\delta
  \phi_i (-p)} + \frac{1}{2} \frac{\delta^2 \Psi_\Lambda}{\delta
  \phi_i (p) \delta \phi_i (-p)} \rb
\end{equation}
$\Psi_\Lambda$ has the correlation functions
\begin{equation}
\vev{\Psi\, \phi_{i_1} (p_1) \cdots \phi_{i_n} (p_n)}^\infty = -
\mu \partial_\mu \vev{\phi_{i_1} (p_1) \cdots \phi_{i_n} (p_n)}^\infty
\end{equation}
Expanding $\Psi_\Lambda$ up to two derivatives, we obtain
\begin{eqnarray}
&&\Psi_\Lambda = 
\int d^2 x\, \Big[ \Lambda^2 \, \dot{a} (\ln \Lambda/\mu; \phi^2/2)
+ \dot{A} (\ln \Lambda/\mu; \phi^2/2)\, (- \partial^2) \frac{\phi^2}{2} \nn\\
&&\qquad + \dot{B} (\ln \Lambda/\mu; \phi^2/2)\, 
\phi_i (- \partial^2) \phi_i  \Big] + \cdots
\end{eqnarray}
where
\begin{equation}
\lb\begin{array}{c@{~\equiv~}l}
\dot{a} (\ln \Lambda/\mu; x) & \frac{\partial}{\partial \ln
  \Lambda/\mu} a (\ln \Lambda/\mu; x)\\
\dot{A} (\ln \Lambda/\mu; x) & \frac{\partial}{\partial \ln
  \Lambda/\mu} A (\ln \Lambda/\mu; x)\\
\dot{B} (\ln \Lambda/\mu; x) & \frac{\partial}{\partial \ln
  \Lambda/\mu} B (\ln \Lambda/\mu; x)
\end{array}\right.
\end{equation}
Especially at $\Lambda = \mu$, the coefficient of
$(\phi^2/2)(- \partial^2) (\phi^2/2)$ is
\begin{equation}
\partial_x \dot{A} (0;x)\Big|_{x=0}
\end{equation}
and that of $\phi_i (-\partial^2) \phi_i$ is
\begin{equation}
\dot{B} (0;0)
\end{equation}

Since $\Psi_\Lambda$ is an infinitesimal change of the Wilson action,
it has two degrees of freedom, corresponding to the infinitesimal
variation of $g$ and that of the normalization of $\phi_i$.
Thus, we can construct two composite operators:
\begin{enumerate}
\item $\Op_g$ that generates an infinitesimal change of $g$:
\begin{equation}
\Op_g \equiv - \partial_g \SL 
\end{equation}
The correlation functions of $\Op_g$ are given by
\begin{equation}
\vev{\Op_g\, \phi_{i_1} (p_1) \cdots \phi_{i_n} (p_n)}^\infty =
- \partial_g \vev{\phi_{i_1} (p_1) \cdots \phi_{i_n} (p_n)}^\infty
\end{equation}
\item $\N$ that generates an infinitesimal renormalization of $\phi_i$:
\begin{eqnarray}
\N &\equiv& - \int_p \phi_i (p) \frac{\delta \SL}{\delta
  \phi_i (p)} - \int_p \frac{\K{p}(1-\K{p})}{p^2} \nn\\
&&\quad \times\lb \frac{\delta \SL}{\delta
  \phi_i (p)} \frac{\delta \SL}{\delta \phi_i (-p)} +
 \frac{\delta^2 \SL}{\delta \phi_i (p) \delta \phi_i (-p)} \rb
\end{eqnarray}
$\N$ counts the number of fields:
\begin{equation}
\vev{\N\, \phi_{i_1} (p_1) \cdots \phi_{i_n} (p_n)}^\infty =
n \vev{\phi_{i_1} (p_1) \cdots \phi_{i_n} (p_n)}^\infty 
\end{equation}
\end{enumerate}
$\Psi_\Lambda$ must be a linear combination of $\Op_g$ and $\N$; hence
\begin{equation}
\Psi_\Lambda = \beta (g) \,\Op_g + \gamma (g) \,\N
\end{equation}
where neither $\beta (g)$ nor $\gamma (g)$ depends on $\Lambda$.  This
gives the differential equation
\begin{equation}
\left( - \mu \partial_\mu + \beta \partial_g \right) \vev{\phi_{i_1}
  (p_1) \cdots \phi_{i_n} (p_n)}^\infty = n \gamma \vev{\phi_{i_1}
  (p_1) \cdots \phi_{i_n} (p_n)}^\infty
\end{equation}
Hence, $\beta (g)$ is the beta function of $g$, and $\gamma (g)$ is
the anomalous dimension of $\phi_i$.

\subsection{$\Op_g$}

The derivative expansion gives
\begin{eqnarray}
\Op_g &=& 
\int d^2 x\, \Big[ \Lambda^2 (- \partial_g)\, a (\ln \Lambda/\mu; \phi^2/2)
+ (- \partial_g) A(\ln \Lambda/\mu; \phi^2/2)\,( - \partial^2)
\frac{\phi^2}{2}\nn\\
&& + (- \partial_g) B (\ln \Lambda/\mu;
 \phi^2/2)\, \phi_i (-\partial^2) \phi_i \Big]
+ \cdots
\end{eqnarray}
At $\Lambda = \mu$, the coefficient of $(\phi^2/2) (-\partial^2)
(\phi^2/2)$ is
\begin{equation}
- \partial_g \partial_x A (0; x)\Big|_{x=0}  = \frac{1}{2} 
\label{Og1}
\end{equation}
and the coefficient of $(1/2) \phi_i (-\partial^2) \phi_i$ is
\begin{equation}
- \partial_g B(0; 0) = 0
\label{Og2}
\end{equation}
These are consequences of the conventions (\ref{normA1}, \ref{normB0}).

\subsection{$\N$}

Using the interaction part of the action, we can rewrite
\begin{eqnarray}
    &&\N = \int_p p^2 \phi_i (p) \phi_i (-p) + \int_p \lb -1 + 2
    \left(1 - \K{p}\right) \rb \phi_i (p) \frac{\delta \SIL}{\delta
      \phi_i (p)}\nn\\ 
    &&\quad - \int_p \frac{\K{p} \left( 1 - \K{p} \right)}{p^2} \lb
    \frac{\delta \SIL}{\delta \phi_i (p)} \frac{\delta \SIL}{\delta \phi_i
      (-p)} + \frac{\delta^2 \SIL}{\delta \phi_i (p) \delta \phi_i (-p)}
    \rb
\end{eqnarray}
Hence, the derivative expansion gives
\begin{eqnarray}
\N (\Lambda) &=& \int d^2
x\, \Big[ \Lambda^2 a_{\N} \left(\ln \Lambda/\mu; \phi^2/2\right)
+ A_{\N} \left(\ln \Lambda/\mu; \phi^2/2\right)\,
\left(- \partial^2\right) \frac{\phi^2}{2} \nn\\ 
&&\quad  + B_{\N} \left(\ln \Lambda/\mu; \phi^2/2\right) \phi_i
\left(- \partial^2\right) \phi_i  \Big] + \cdots
\end{eqnarray}
where
\begin{eqnarray}
A_{\N} \left(\ln \Lambda/\mu; x\right) &=& - 2 A\left(\ln
    \Lambda/\mu;x\right) - 2 x \frac{\partial}{\partial x} A \left(\ln
    \Lambda/\mu; x\right)  + \cdots\\ 
B_{\N} \left(\ln \Lambda/\mu; x\right) &=& 1 - 2 B \left(\ln
    \Lambda/\mu; x\right) + \cdots 
\end{eqnarray}
up to loop corrections.  At $\Lambda = \mu$, the coefficient of
$(\phi^2/2) (-\partial^2) (\phi^2/2)$ is
\begin{equation}
\partial_x A_{\N} (0; x)\Big|_{x=0} = 2 g + \cdots
\label{N1}
\end{equation}
and the coefficient of $(1/2) \phi_i (-\partial^2) \phi_i$ is
\begin{equation}
B_{\N} (0; 0) = 1 + \cdots
\label{N2}
\end{equation}

\subsection{$\beta$ and $\gamma$}

Comparing the derivative expansion of $\Psi_\Lambda$ with those of
$\Op_g$ and $\N$, we obtain $\beta (g)$ and $\gamma (g)$ as follows:
\begin{eqnarray}
\partial_x \dot{A} (0;x)\Big|_{x=0} &=& \frac{1}{2} \beta (g) + \gamma (g)
\partial_x A_{\N} (0;x)\Big|_{x=0}\label{Adot}\\
\dot{B} (0;0) &=& \gamma(g) B_{\N} (0;0)\label{Bdot}
\end{eqnarray}
where we have used (\ref{Og1}, \ref{Og2}).

Using (\ref{A11}, \ref{B01}), we obtain
\begin{equation}
\partial_x \dot{A}^{(1)} (0;x)\Big|_{x=0} = N \frac{g^2}{4\pi},\quad
\dot{B}^{(1)} (0;0) = \frac{g}{4\pi}
\end{equation}
Using (\ref{N1}, \ref{N2}), we also obtain
\begin{equation}
\partial_x A_{\N}^{(0)} (0;x)\Big|_{x=0} = 2g,\quad
B_{\N}^{(0)} (0;0) = 1
\end{equation}
Hence, at 1-loop (\ref{Adot}, \ref{Bdot}) give
\begin{eqnarray}
N \frac{g^2}{4 \pi} &=& \frac{1}{2} \beta^{(1)} + \gamma^{(1)} \cdot 2 g\\
\frac{g}{4 \pi} &=& \gamma^{(1)} \cdot 1
\end{eqnarray}
Thus, we obtain (\ref{1loop-betagamma}).


\newpage

\section*{References}

\begin{thebibliography}{99}
\bibitem{Polyakov:1975rr}
  A.~M.~Polyakov,
  Phys.\ Lett.\  B {\bf 59}, 79 (1975).
\bibitem{Bollini:1972ui}
  C.~G.~Bollini and J.~J.~Giambiagi,
  Nuovo Cim.\  B {\bf 12}, 20 (1972).
\bibitem{'tHooft:1972fi}
  G.~'t Hooft and M.~J.~G.~Veltman,
  Nucl.\ Phys.\  B {\bf 44}, 189 (1972).
\bibitem{Wilson:1973jj}
  K.~G.~Wilson and J.~B.~Kogut,
  Phys.\ Rept.\  {\bf 12}, 75 (1974).
\bibitem{Polchinski:1983gv}
  J.~Polchinski,
  Nucl.\ Phys.\  B {\bf 231}, 269 (1984).
\bibitem{Igarashi:2009tj}
  Y.~Igarashi, K.~Itoh, and H.~Sonoda,
  ``Realization of symmetry in the ERG approach to quantum field theory,''
  arXiv:0909.0327 [hep-th].
\bibitem{Becchi:1996an}
  C.~Becchi,
  ``On the construction of renormalized gauge theories using  renormalization
  group techniques,''
  arXiv:hep-th/9607188.
\bibitem{Pernici:1998tp}
  M.~Pernici and M.~Raciti,
  Nucl.\ Phys.\  B {\bf 531}, 560 (1998)
  [arXiv:hep-th/9803212].
\bibitem{Sonoda:2002pb}
  H.~Sonoda,
  Phys.\ Rev.\  D {\bf 67}, 065011 (2003)
  [arXiv:hep-th/0212302].
\bibitem{Sonoda:2006ai}
  H.~Sonoda,
  J.\ Phys.\ A  {\bf 40}, 5733 (2007)
  [arXiv:hep-th/0612294].
\bibitem{'tHooft:1973mm}
  G.~'t Hooft,
  Nucl.\ Phys.\  B {\bf 61}, 455 (1973).
\end{thebibliography}
\end{document}